\documentclass[runningheads]{llncs}
\usepackage{graphicx}
\usepackage{float}
\usepackage{dirtytalk}
\usepackage{amsmath}
\usepackage{amsmath,amssymb}

\usepackage{comment}
\usepackage{xspace}
\usepackage{caption}
\usepackage{subcaption}
\usepackage{amsmath}
\DeclareMathOperator*{\argmax}{arg\,max}

\begin{document}
\title{Investigation of Topic Modelling Methods for Understanding the Reports of the Mining Projects in Queensland}
\titlerunning{Investigation of Topic Modelling: Mining Projects in Queensland}

\author
{
Yasuko Okamoto\inst{1}\and
Thirunavukarasu Balasubramaniam\inst{2}\and
Richi Nayak\inst{2}
}
\authorrunning{Yasuko Okamoto et al.}

\institute{RecordPoint, Australia.\\ \email{yasuko.okamoto0086@gmail.com}  \and
School of Computer Science and Centre for Data Science,
Queensland University of Technology,
Brisbane, Australia. \\ \email{\{thirunavukarasu.balas, r.nayak\}@qut.edu.au}\\
}
\maketitle              % typeset the header of the contribution
\begin{abstract}
In the mining industry, many reports are generated in the project management process. These past documents are a great resource of knowledge for future success. However, it would be a tedious and challenging task to retrieve the necessary information if the documents are unorganized and unstructured. Document clustering is a powerful approach to cope with the problem, and many methods have been introduced in past studies. Nonetheless, there is no silver bullet that can perform the best for any types of documents. Thus, exploratory studies are required to apply the clustering methods for new datasets. In this study, we will investigate multiple topic modelling (TM) methods. The objectives are finding the appropriate approach for the mining project reports using the dataset of the Geological Survey of Queensland, Department of Resources, Queensland Government, and understanding the contents to get the idea of how to organise them. Three TM methods, Latent Dirichlet Allocation (LDA), Nonnegative Matrix Factorization (NMF), and Nonnegative Tensor Factorization (NTF) are compared statistically and qualitatively. After the evaluation, we conclude that the LDA performs the best for the dataset; however, the possibility remains that the other methods could be adopted with some improvements.

\end{abstract}

\section{Introduction}
For the mining industry, documentation is one of the key components. The state governments in Australia publish the framework for the mining project management and instruct the documentation in detail. Thus, a large number of documents are generated as the project milestone products every year. These documents comprise feasibility studies, risk location, change management, etc., and they contain knowledge for future success. Therefore, building the document management and information retrieval system is a crucial task in the industry. Regarding the mining projects in Queensland, the Geological Survey of Queensland (GSQ), Department of Resources, Queensland Government has collected the project reports form mining companies over the past hundred years. The collection is quite complex since it consists of more than 100,000 reports submitted by over 5,000 companies. Furthermore, the documents are currently unstructured and unorganised, which means the documents are not searchable, and information retrieval from such a collection is a tedious task.

Document clustering is a text mining approach to cope with the problem by grouping document based on their topic and similarities by automatically finding patterns and characteristics from data itself without any predefined data labels. With its capability, document clustering has been widely applied to understand the collection of documents and build a robust search engine \cite{gupta2012role}. The paper will focus on one of the document clustering techniques, the topic modelling (TM). TM discovers topics hidden in the collection as well as describes the association between the topics and each document. The advantage is that it does not only group the documents but tells the contents of each group simultaneously. It has been studied by many researchers, and the past research has developed multiple algorithms, such as probabilistic latent semantic analysis (PLSA) \cite{hofmann1999probabilistic}, latent Dirichlet allocation (LDA) \cite{blei2003latent}, and non-negative matrix factorization (NMF) \cite{xu2003document}. In recent years, the studies on the application of TM methods have been also active, especially for the domain where the expertise is required, for example, bio-informatic, healthcare, and medical \cite{huang2011enhanced,dantu2021exploratory,feng2020topic,balasubramaniam2021identifying}.

While a variety of methods and use cases have been introduced, the nature of text mining is domain-specific. It means that the appropriate approach can vary depending on the data types and purposes. Therefore, the exploratory studies are continuously needed for the application in the real-world. Thus, this study investigates the performance of several TM methods and explores the best approach for GSQ’s documents. Besides, we expect that the grouping results obtained from the experiment would be a help to understand the contents of the collection and analyse the trends of project and business. Furthermore, these findings could derive the category labels and the summary of the contents of each group, which can be essential factors and materials for future studies on building a robust document management system for GSQ.

The rest of this paper is structured as follows. Section 2 elaborates the recent studies on TM methods and applications and identifies the problem space to be addressed in the study. Section 3 describes the proposed methods. Algorithms, datasets, and evaluation methods applied in the experiment will be explained. The results of TM will be observed and analysed in section 4. Finally, section 5 provides the conclusion and suggestions for future studies.

\section{Related Works}
\subsection{Document representation model}
The basic concept of TM methods is taking the numeric (document x term) vectors that represent the documents as input and converting them into (topic x term) vectors and (document x topic) vectors. How documents should be represented in the vector space, or the document representation model is one of the important parameters for TM methods. This sub-section briefly reviews and introduces several models. The most classic one is Bag-of-words (BoW) model where documents are represented as the collection of words, and the word order information in the document is ignored. In the simple word count vector, for example, a document is a sequence of $N$ words denoted by $D = \{tf_1, tf_2, ... , tf_N\}$, where $tf_i$ is the frequency of term $i$ in document $D$. The term-frequency-inversed-document-frequency (TFIDF) is a popular scheme of the BoW model. It adds the importance of the term in the collection to the count vector. However, there was a common awareness of the limitation of these approaches among the researchers: the document representation models lose the semantics and relationships among words. The powerful technique to address this problem is word-embedding. In 2013, the research team at Google advocated a new approach, called Word2Vec \cite{mikolov2013efficient}, which involved two-layers neural networks and embedded each word to a numeric vector space. The technique allows to calculate the semantic similarity between words, and a document can be represented by a vector incorporating the relationship between the words using the average values of Word2Vec \cite{chen2017efficient}. The technique was further developed and expanded to the document scale, which is called Doc2Vec \cite{le2014distributed}.

\subsection{Recent Studies of Topic Modelling}
Many TM methods have been developed in past studies, and they can be categorised into two types: the probabilistic model and the non-probabilistic model. The examples of former are PLSA \cite{hofmann1999probabilistic} and LDA \cite{blei2003latent}. Both calculate the probability of the word appearing in the different topics, P(word|topic), and the probability of the topics in the document P(topic|document). Latent Semantic Analysis (LSA) \cite{deerwester1990indexing} and NMF \cite{xu2003document} are categorised in non-probabilistic approaches that are algebraic approaches using matrix factorization. While there are several methods as above mentioned, LDA and NMF are the most popular techniques for industrial application \cite{westerlund2018topic,zhang2019forty,moro2019text}. %More research on the two techniques are found in the last decade. TABLE I. presents several related works for applying TM to analyse the contents of domain-specific long documents.

At the same time, researchers keep investigating better TM approaches using different datasets. Studies of Anantharaman, et al., Ray, et al., and Chehal, et al. compared the existing methods explained above, and the conclusions of the studies show that the best methods are different depending on the datasets \cite{anantharaman2019performance,ray2019review,chehal2021implementation}. Furthermore, the researchers sometimes modified the exiting methods to adjust them according to the datasets or purposes. Wang, et al. introduced the Multi-Attribute LDA that can consider additional document attributes other than terms when mining the topics \cite{wang2015detecting}. With the extended LDA, they concluded that hot topics in the microblog could be found more properly using not only texts, but the post time and hashtag information compared to the normal LDA. Shi, et al. proposed a semantics-assisted non-negative matrix factorization (SeaNMF) \cite{shi2018short}. While the original NMF uses the BoW document model, they argued that incorporating the semantic correlations between words discovered more coherent topics compared with other methods such as normal LDA and NMF. %The recent studies on the investigation of multiple TM methods are summarised in TABLE II.

The recent studies imply that the best approach varies depending on the datasets, and the comparative studies are required when applying TM to the new datasets while LDA and NMF are the most popular methods. Moreover, the methods sometimes should be modified according to the purpose. Therefore, this research investigates multiple methods to answer What TM approach is appropriate to analyse the contents of GSQ’s mining project reports?

\section{Topic Modelling Approaches}
Given a large number of documents and companies who submitted the reports, the study focuses on two specific objectives: grouping documents by contents similarities, and grouping companies based on their reports. For the former objective, LDA and NMF are applied considering their popularities and characteristics that worked better on the long texts as seen in the recent studies. To achieve the latter objective, 3-dimensional TM that can take the company information of documents into account.

\subsection{Two-dimensional Topic Modelling: LDA and NMF}
Both algorithms accept the input of (document x term) matrix and generate two matrices that are (document x topic) matrix and (topic x term) matrix. The former denotes how each document is associated with each topic, and the latter can be considered as topics list discovered from the collection. In the study, each document is assigned to one topic group using the (document x topic) matrix. More precisely, document $i$ is assigned to topic $x$ if $x$ = $\argmax$$_j$ $v_{ij}$, where $v_{ij}$ each element of (document x topic) matrix representing document $i$ is associated with topic $j$. This assigning method is referred to as the argmax method in the rest of the paper.

LDA is a probabilistic model introduced by Blei et al. \cite{blei2003latent}, and the basic concept is assuming that each document consists of multiple topics, where each topic is a probability distribution over words. Assume $K$ is the number of topics, $N$ is the number of words in the document, $M$ is the number of documents in the collection, $\alpha$ is the parameter representing the Dirichlet prior for the per-document topic distribution, $\beta$ is the parameter representing the Dirichlet for the per-topic word distribution, $\varphi(k)$ is the word distribution for topic $k$,$\theta(i)$ is the topic distribution for document $i$, $z(i,j)$ is the topic assignment for $w(i,j)$, $w(i,j)$  is the word $j$ in document $i$. The aim is to learn  $\varphi$ (topic x term) matrix, and $\theta$ (document x topic) matrix. $\alpha$, $\beta$, and $K$ should be specified by the user. In this study, exploring the best K is focused on. Therefore, $\alpha$ and $\beta$ are set to the default values of LDA function of scikit-learn package, which are 1/$K$. The range of $K$ is explained in the coming section of the paper.

NMF is a matrix factorization method. Given the original matrix $\boldsymbol{\mathrm{X}}$ (document x term), two matrices can be obtained $\boldsymbol{\mathrm{W}}$ (topic x term) and $\boldsymbol{\mathrm{H}}$ (topic x document), such that $\boldsymbol{\mathrm{X}}$ = $\boldsymbol{\mathrm{WH}}$. By taking advantage of the fact that $\boldsymbol{\mathrm{X}}$ is non-negative, the two matrices $\boldsymbol{\mathrm{W}}$ and $\boldsymbol{\mathrm{H}}$ are optimised over the following objective function using Frobenius norm:

\begin{equation}
    \frac{1}{2} \|\boldsymbol{\mathrm{X}} - \boldsymbol{\mathrm{W}}\boldsymbol{\mathrm{H}}\|^2 = \sum_{i=1}^{n}\sum_{j=1}^m(\boldsymbol{\mathrm{X}}_{ij}-(\boldsymbol{\mathrm{WH}}_{ij}))^2
\end{equation}

In the function, the error of reconstruction between $\boldsymbol{\mathrm{X}}$ and the product of $\boldsymbol{\mathrm{W}}$ and $\boldsymbol{\mathrm{H}}$ is measured. Thus, $\boldsymbol{\mathrm{W}}$ and $\boldsymbol{\mathrm{H}}$ are updated iteratively using the following update function until convergence.

\begin{equation}
    \boldsymbol{\mathrm{W}} \xleftarrow{} \boldsymbol{\mathrm{W}}\frac{\boldsymbol{\mathrm{XH}}}{\boldsymbol{\mathrm{WHH}}}
\end{equation}

\begin{equation}
    \boldsymbol{\mathrm{H}} \xleftarrow{} \boldsymbol{\mathrm{H}}\frac{\boldsymbol{\mathrm{WX}}}{\boldsymbol{\mathrm{WWH}}}
\end{equation}

In the study, only $K$ is tuned using multiple values. The other parameters other than the method to initialise the procedure were set to the default values of NMF function of scikit-learn package. For the initialisation method, Nonnegative Double Singular Value Decomposition (NNDSVD) is used since it is considered better for sparse data \cite{boutsidis2008svd}.

\subsection{Three-dimensional (3D) Topic Modelling: Tensor clustering}
3D TM can accept a 3D matrix while LDA and NMF accept a 2D matrix. Hence, it can group data with two criteria simultaneously. In the study, a 3D matrix (document x company x term) matrix is fed into the algorism, and it generates three 2D matrices that are (document x topic) matrix, (term x topic) matrix and (company x topic) matrix. The third matrix describes how each company is associated with each topic. Thus, the method enables to group documents and group companies at once. The method is referred to as the tensor clustering in the paper. The applied method in this study is the Saturating Coordinate Descent (SaCD) that is an improved method of Nonnegative Tensor Factorization (NTF) \cite{balasubramaniam2020efficient}. The NTF model is formulated as follows:

\begin{equation}
\min_{\boldsymbol{\mathrm{U,V,W}} \geq 0} f(\boldsymbol{\mathrm{U,V,W}}) = \|\boldsymbol{\mathcal{X}} - \boldsymbol{\mathrm{U,V,W}}\|^2
\end{equation}

where $\boldsymbol{\mathcal{X}}$ is a 3D tensor, which is (document $\times$ term $\times$ company) tensor here, and $\boldsymbol{\mathrm{U,V,W}}$ are the factor matrices for $\boldsymbol{\mathcal{X}}$. SaCD makes this algorithm faster and more scalable applying the updated method that reduces the calculation in the learning process. After obtaining (document $\times$ topic) matrix and (company $\times$ topic) matrix, each document and each company is assigned to one topic using the argmax method in the same manner as 2D TM.

\subsection{Dataset}
This section discusses the datasets used in the experiment. We use only the reports that had been converted into Microsoft (MS) Word format and stored. Besides, the report can consist of more than one documents. In that case, only the body of the report is used.% assuming that the document file whose name ends up it with "_001.docx" is the body.

%The actual target datasets for evaluation are selected based on the statistics about GSQ's reports. The statistics are obtained from the metadata, and the mining industry categories of the reports refer to the category file "metadata_original.xlsx". Both files are provided by the project supervisory team.

Given that the collection is quite large, and categorised by mining industries, such as coal, mineral, petroleum, it is assumed that performing TM methods onto the subsets of each category would provide new findings rather than performing TM onto the whole collection. Therefore, this study focus on performing TM methods on the target subsets. The target subsets are chosen by the following steps: According to the metadata file, more than 60\% of the reports were submitted in the last two decades. Therefore, the target period is set to 2000 – 2020. Next, by aggregating the reports of this period, the three dominants categorise are found: Coalbed Methane, Gold and Coal. In this paper we only report the outputs of Coal. %To be more precise, each document can be categories in one or more categories (e.g. Report No. 100804 belongs to Coalbed Methane and Gas), so the category “Coalbed Methane” in the paper refers to Coalbed Methane or Coalbed Methane and other categories.

%Same applies to Gold and Coal. Thus, the three categories are set to the targets in the experiment.

Moreover, the reports of each category have the information about report types, such as well completion report, annual report, and final report. Hence, the target was further narrowed down by report type. In the coal, the annual report is the main report type that occupies 80\%. 

After setting the target period, categories, and report types, the number of reports available in MS word format is counted. The data is also cleaned by the company name identification and removing the documents that had not been converted into MS word format properly. 

\subsection{Data Preprocessing}
\subsubsection{Text Extraction}
The process was to parse the reports in Microsoft word format and extracting the body text with python programming. Tab and newline were removed here. At the end of the process, the data frame that contains report – body text (a string) information is obtained.

\subsubsection{Text Preprocessing}
The text pre-processing is the process to clean the documents using the natural language processing (NLP) methods before converting them into the numeric data. The details are as follows:

\begin{enumerate}
    \item Tokenisation: In the tokenisation process, the texts in documents are split text into words and lowercased using spaCy, the NLP package in Python. Punctuations, special characters, words with less than three characters are removed after tokenisation.
    \item Stop-word removal: Stop-words are the terms that have little meaning and occur in the document with high frequencies, such as delimiters and prepositions. The terms are removed before performing TM. The stop-words list from the NLTK corpus is applied. Besides, the additional 13 stop-words are chosen, which are: appendix, area, Australia, fax, figure, ltd, map, page, phone, project, report, year, within.
    \item Stemming: Stemming is the process to uniform the words in the different morphological forms. For instance, the term "program" can be different forms such as "programs", "programmer", "programmers", "programming " will be uniformed into “program. The Porter Stemming Algorithm advocated by Martin Porter in 1980 is applied using the NLTK package.
\end{enumerate}

\subsubsection{Data Transformation}
The documents transformation process is to convert the documents list into a numeric matrix to feed the TM algorithm. As discussed in section 2, there are several document representation models; however, following the past studies that introduced LDA and NMF for TM \cite{blei2003latent,xu2003document}, the BoW model is adopted. More specifically, each document is vectorized with the term-frequency (TF) for LDA and TFIDF values for NMF. As explained briefly in section 2, TFIDF is the weighting schema that adds the importance of the term in the collection.

All documents in the dataset are converted into vectors in the same manner. Thus, the 312 documents of Coal dataset with 8,549 unique terms, are converted into 2D TF matrix and TFIDF matrix with the shape of (312, 8,549) for LDA and NMF. For the 3D tensor clustering, TF values are adopted, and the company information is added to the (document $\times$ term) TF matrix. Therefore, the Coal dataset with 76 companies is converted into the 3D matrix with the shape of (312, 76, 8549) for the tensor clustering. 

\subsection{Evaluation measures}
Silhouette analysis is one of the major evaluation methods of clustering approach and can be used to analyse the separation distance between the clusters. In the method, the silhouette coefficient of each sample is calculated, which indicate how much the sample is far away from the neighbouring clusters with the following equation.

\begin{equation}
    S_i = (b_i - a_i)/max(a_i,b_i)
\end{equation}

where $S_i$ the silhouette coefficient of sample $i$, $a_i$ the average distance between $i$ and all the other data points in the cluster to which $i$ belongs, and $b_i$ is the minimum average distance from $i$ to all clusters to which $i$ does not belong. The value of the silhouette coefficient is between [-1, 1]. The value closer to 1 denotes that the sample is far from the neighbouring clusters,and the negative value implies the sample might have been assigned to the wrong cluster. In the study, silhouette scores are calculated using (topic x document) matrix for LDA/NMF. For the tensor clustering, (topic x document) matrix is used to evaluate the grouping quality for document groups, and (topic x company) matrix is used for the quality of company groups. For each $K$, the average silhouette score is calculated, as well as the silhouette coefficients of all samples are plotted in the figure to observe whether the samples of each group were assigned properly. The candidate numbers for $K$ are chosen for each dataset based on the average silhouette scores. Then, the plotting results of silhouette scores and the results of the second evaluation method, topic keywords matching, are taken into account to decide the best number for $K$ among the candidates and the best TM methods for grouping documents.

\subsection{Topic Keyword Matching}
After selecting candidate $K$s with the above method, for each $K$, the top-30 keywords of each group are observed. The top-30 keywords are obtained from (topic $\times$ term) matrix generated by TM. The 30 words are compared with the 30 terms appearing in the documents of the group most frequently. The purpose of the
method is to check whether the keywords found by TM actually appear in the documents of the group. %For example, TABLE X. shows that 27 keywords out of 30 (the words in orange in the table) for topic \#0 of Coalbed Methane dataset found by LDA match the top-30 frequent terms appearing in the documents belonging to topic \#0. Thus, the matching ratio becomes 90\%.

\section{Results and Discussion}
The purpose of 2D TM is grouping documents. The best K and the best method were identified among the results of LDA,NMF, and the grouping documents result of the tensor clustering.

\begin{figure}
    \centering
    \includegraphics[width = 3in, height = 2in]{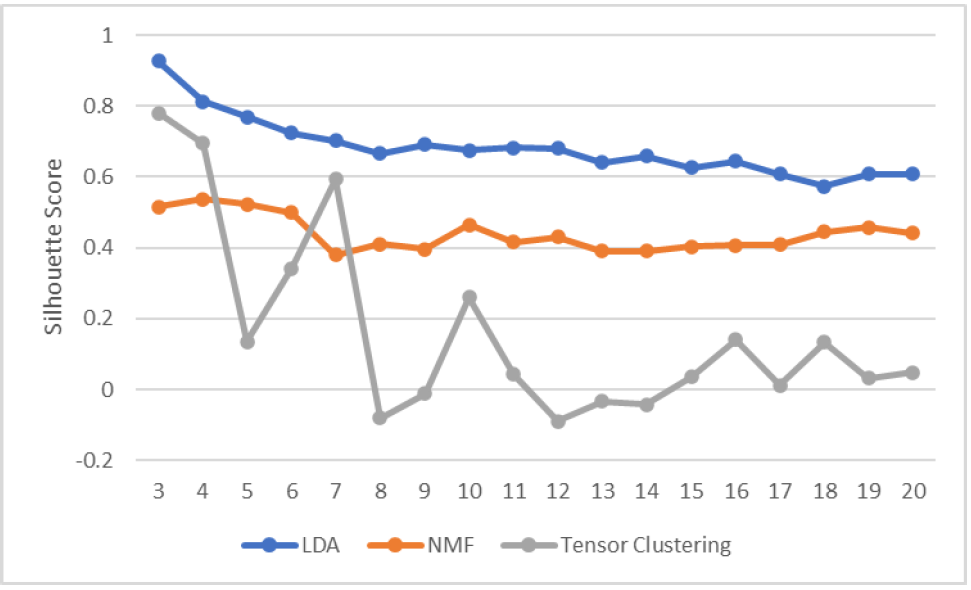}
    \caption{Silhouette score comparison}
    \label{fig:sil_comp}
\end{figure}

\begin{figure}
    \centering
    \includegraphics[width = 3in, height = 2in]{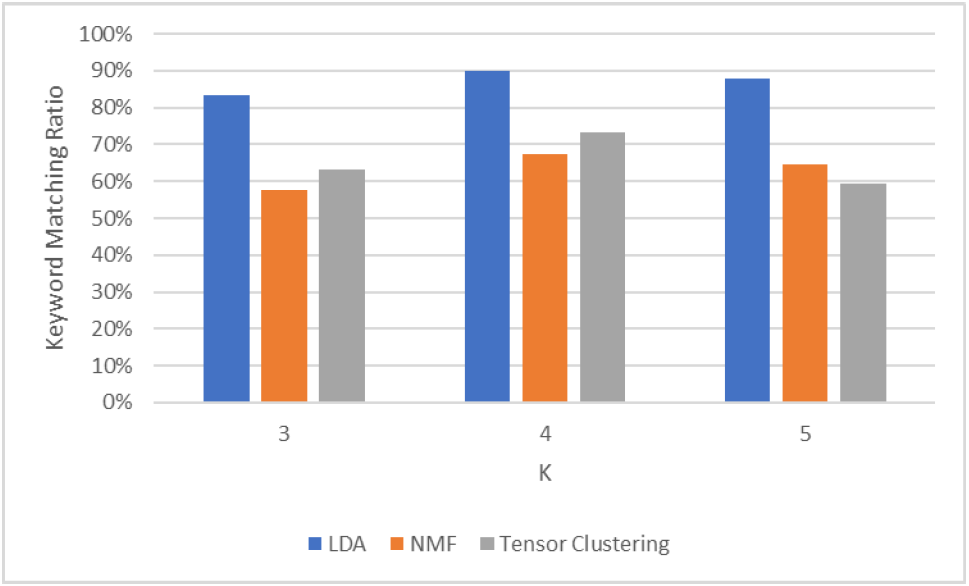}
    \caption{Keyword matching ratio comparison with candidate number of K}
    \label{fig:word_comp}
\end{figure}

For the Coal dataset, K = 3 and 4 resulted in the highest average silhouette scores for all algorithms (Figure \ref{fig:sil_comp}). Furthermore, the silhouette score comparison and the keyword matching ratio comparison (Figure \ref{fig:word_comp}) shows that LDA performed the best. Finally, from the facts that the keyword matching ratio is higher, and the silhouette analysis resulted better when K = 4, 4 was chosen as the best number of groups for Coal dataset.

With 3D TM, we aimed to group documents and companies simultaneously. Therefore, the tensor clustering method was evaluated separately from 2D TM results. Overall, it can be mentioned that there are some doubts about the results of grouping companies for all dataset because the detailed silhouette analysis indicates most of the companies are assigned into the same group, and the companies that were assigned into the different groups resulted in the negative silhouette score.

For Coal dataset, 3 and 4 could be candidates for $K$ according to the silhouette score (Figure Figure \ref{fig:sil_comp}). Observing the keyword match ratio, the better result is provided when $K$ = 4, which was 73\%. Hence, 4 was chosen as the best K. However, it should be noted that the detailed silhouette analysis for both grouping document and grouping companies indicate that the qualities of some clusters are low, which means that the tensor clustering could not perform well for this dataset.

From the evaluation results, the best $K$ and method were chosen for each dataset as below. The following sub-sections discuss the details of grouping results of chosen $K$ and method.

\subsection{Document group by LDA}
The dataset containing 312 documents submitted by 76 companies were divided by LDA into four groups: 25 documents of 6 companies in topic \#0, 253 documents of 69 companies in topic \#1, 21 documents of 3 companies in topic \#2, 13 documents of 8 companies of topic \#3. What each topic represents is as follows: Topic \#0 is about “volcan”, “creek”, “seismic”, “cockenzi”, “permian”, topic \#1 is about “explor”, “seam”, “basin”, “measur”, “mine”, topic \#2 is about “basin”, “moreton”, “walloon”, “surat”, “basal”, topic \#3 is about “grey”, “fresh”, “grain”, “medium”, “siltston”. Most of or all of the document of the same company were assigned to a single topic. Therefore, it can be considered the contents of reports submitted by the same company are consistent in the dataset, or each company can have a strong relationship with a single topic.

\subsection{Document group by Tensor}
Four topics found by the tensor clustering are: topic \#0 about “sandston”, “siltston”, “medium”, “grey”, “grain”, topic \#1 about “epc”, “rock”, “medium”, “sandston”, “seam”, topic \#2 about “thick”, “grain”, “grey”, “sandston”, “dark”, topic \#3 about “rock”, “grain”, “grey”. “sandston”, “clay”. As seen from these keywords, the four topics are similar. However, LDA results showed that there should be several topics in the dataset. Therefore, by removing some common words such as “coal”, “drill”, and “geolog” as stop-words, the method could find more meaningful topics. Company groups and document groups are consistent. In both grouping results, topic \#0 is occupied by Tenement Administration Services, topic \#1 is occupied by Lance Grimstone \& Associates Pty Ltd, topic \#2 is occupied by BHP Billion Mitsubishi Alliance (BMA). The rest companies with 271 documents were assigned to topic \#1. Thus, from the tensor clustering, the three companies that have unique contents in their reports were found.

\subsection{Indecisiveness of Tensor Clustering}
As discussed above, the statistical evaluation score of tensor clustering was low, and inconsistent results were observed.% in Coalbed Methane and Gold dataset. 
The reason for the low clustering quality was investigated under the assumption that the tensor clustering is more indecisive and the argmax method would not be appropriate. TM is originally a soft clustering technique, which does assess the association of each document with each topic. Thus, a document can belong to one or more groups. The argmax method, however, assigns each document to one topic that has the strongest association with the document. In the process, the relationships that the document might have with other topics are ignored. Hence, this method is not appropriate if the TM method is softer or more indecisive since the impact of ignored relationships would be large. For the investigation, the standard deviations of (document x topic) matrix and (company x topic) matrix are calculated and compared among TM methods. %To be more precise, TABLE XXI. is an example of (company x topic) matrix, and each cell denotes the association score of a document with each topic. Using the matrix, standard deviations of each row is calculated, and then the average of all companies is obtained. 
In the same manner, the average standard deviations are obtained from (document x topic) matrices of all TM methods. If the standard deviation is large, the method is evaluated to be decisive. On the other hand, the method is indecisive if the value is small.

\begin{figure}
    \centering
    \includegraphics[width = 3in, height = 2in]{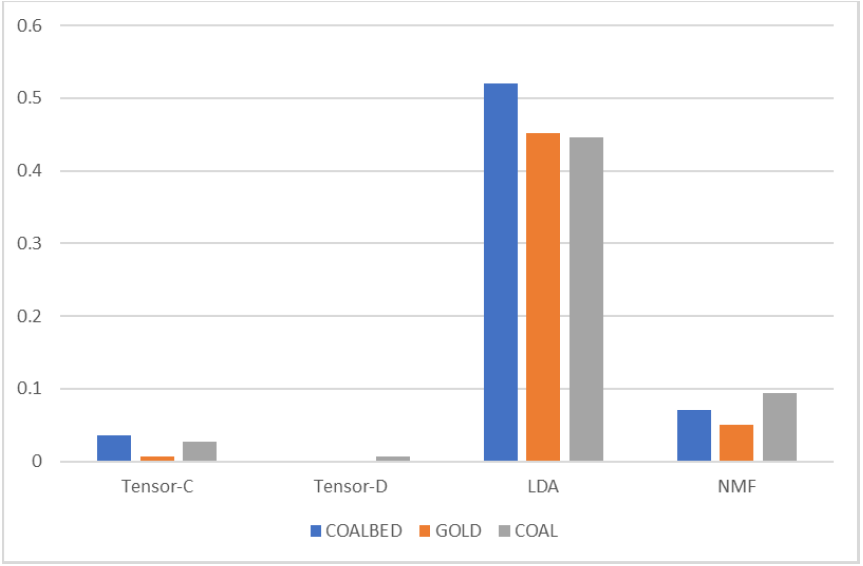}
    \caption{Comparison of Standard Deviation}
    \label{fig:std}
\end{figure}

Comparing the standard deviations from the matrices of the best $K$ (Figure \ref{fig:std}), the standard deviation of the tensor clustering is smaller than LDA and NMF in all dataset. On the other hand, the value of LDA is much larger compared to other methods. It corresponds to that LDA resulted in the highest silhouette analysis score as seen in the evaluation section.

Furthermore, by comparing the standard deviations of all $K$s, it is found that LDA is the most decisive, and the tensor clustering is the least decisive. Thus, it can be concluded that the argmax method is not appropriate for the tensor clustering, and the other methods need to be explored to use the outputs of tensor clustering more effectively.

\section{Conclusion}
Comparing multiple TM methods for GSQ’s reports, we conclude that LDA performs the best. However, at the same time, the need for improvement remains since it found many overlapping words over topics. For improvement, we can consider the modification of text pre-processing. Especially, some new stop-words should be added in the future experiment. %We suggest the additional stop-words in the Appendix.

Although LDA performed the best in term of the quality of clusters, testing the other methods is still meaningful. This study involved a small dataset, but we have to expand our methods to the whole GSQ’s collection in the future. Hence, faster and scalable algorithms would be more ideal. The tensor clustering method could respond to this need as it can perform much faster than LDA. Furthermore, it enables us to analyse the document from the various aspects simultaneously. However, to adopt this algorithm, it has to be improved. For the future study, we suggest the following things as the idea for the improvement: First, exploring the new assigning method other than the argmax is required. Implementing other clustering methods, such as K-Means and DBSCAN, on the output of the tensor clustering can be options. Second, the new stop-words should be added in the text pre-processing. Third, normalising the number of documents per companies can be considered.

Finally, some characteristics from dataset were found by observing the results of the best $K$. However, in the document clustering approach, the statistical evaluation score does not necessarily mean that the grouping results are comprehensive for human. Hence, the human interpretation of clustering results is still required. In the study, we recorded and visualised the results of all $K$s using a BI tool, Tableau. We would suggest reviewing the results of other $K$s in the future if the results of the best $K$ are considered unreasonable by the experts in the industry.

\section*{Acknowledgement}
This research was supported by the Geoscience Information, Geological Survey of Queensland (GSQ), Department of Resources, Queensland Government.
\bibliographystyle{splncs03_unsrt}
\bibliography{air.bib}

\end{document}